\documentclass[english,preprint,aps,pre,preprintnumbers,floatfix,showpacs]{revtex4-1}
\usepackage{epsfig}
\usepackage{float}
\usepackage{float}
\usepackage{float}
\usepackage{graphicx}

\begin{document}

\title{Huge broadening of the crystal-fluid interface
for sedimenting colloids}
\author{Elshad Allahyarov$^{1,2,3}$, Hartmut L{\"o}wen$^1$}
\affiliation{ 
{(1)} Institut f{\"u}r Theoretische Physik,
Heinrich-Heine-Universit{\"a}t D{\"u}sseldorf, D-40225 D{\"u}sseldorf,
Germany \\
{(2)} Joint Institute for High Temperatures, Russian Academy of Sciences,
 Izhorskaya 13/19 , 117419 Moscow, Russia\\
{(3)} Department of Physics, Case Western Reserve University, Cleveland, 
44106 Ohio, USA 
 }

\begin{abstract}
For sedimenting colloidal hard spheres, 
the propagation and broadening of the crystal-fluid interface is studied
by Brownian dynamics computer simulations of an initially homogeneous sample.
Two different types of interface broadenings are observed: the first occurs
during growth and is correlated with the interface velocity, 
the second is concomitant with the splitting of the 
crystal-fluid interface 
{ into the crystal-amorphous and amorphous-liquid interfaces}. 
The latter width is strongly peaked as a function of the gravitational 
driving strength with a huge amplitude relative to its equilibrium counterpart.
\end{abstract}
\maketitle

\section {\bf Introduction}
Sedimentation of colloidal particles in a liquid is a widespread phenomenon
which governs the formation of river sediments close to salty sea water \cite{river},
controls the function of red blood cells \cite{red_blood_cell},
and is technologically used to separate different sorts of particles
by centrifugation \cite{Philipse}. Gravity is also used to compactify colloidal samples.
In a finite container, the actual settling process reaches a
sedimentation-diffusion equilibrium, which is characterized by a static
colloidal density profile. Much of our particle-resolved knowledge of
dense suspensions stems from model hard sphere colloidal dispersions 
made up of sterically-stabilized particles. In equilibrium, it has been shown that
density profiles contain the isothermal equation of state of the hard sphere system
\cite{Piazza,Barrat}.
At high gravitational strengths (or Peclet numbers) 
{ the density of hard spheres at the bottom of suspension increases above 
the lower limit of the solid-liquid coexistence $\phi=0.492$ \cite{Davidchack}, and}
a crystallization in the  bottom layers occurs \cite{Biben,Marechal,Mori2}. 
The number of crystalline layers is controlled by gravity
{ and by  the  sedimentation velocity of particles,}
 and the overall particle density per area.
While the equilibrium properties of crystal sediments are well-explored by now,
the dynamics and relaxation towards equilibrium is by far less understood 
\cite{Schmidt_Dzubiella,Soft_Matter}.
In particular, it is known that crystallization occurs under gravity at different 
conditions than in the bulk \cite{Ackerson,Cheng}. 
The technique of colloidal templating has been used to steer
colloidal crystallization layer-by-layer in gravity
\cite{Hoogenboom2,Ramsteiner} and the kinetics of crystalline 
defects \cite{Mori,Hoogenboom1} has been debated for sediments.

\begin{figure}  [!ht]
\begin{center}
\includegraphics*[width=0.9\textwidth]{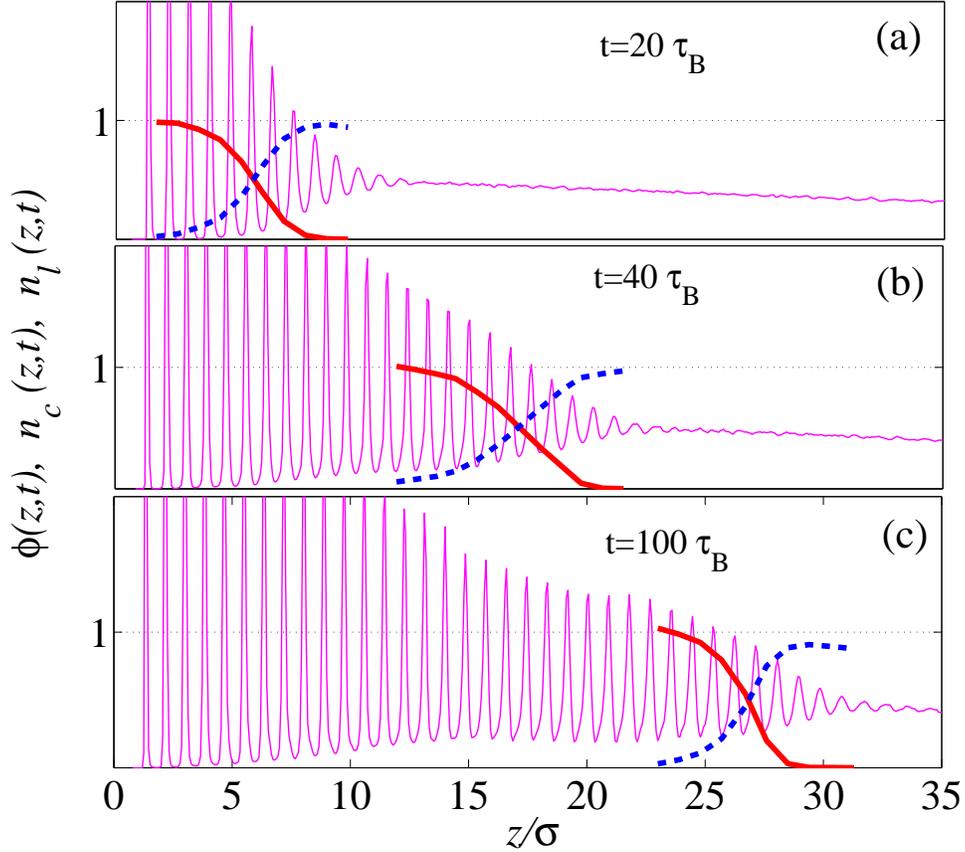}
\end{center}
\caption{Laterally averaged packing fractions $\phi(z,t)$ (thin pink lines)
 versus reduced height $z/\sigma$
 for three different  times (a-c): $t=20\tau_B$,$40\tau_B$,$100\tau_B$.
The parameters are $Pe=0.5$,  $\eta_{A}=45.4$, and $\phi$=0.3. 
Also shown  are the interfacial profiles $n_{c}(z,t)$ for 
the "crystalline" particles  (thick red line)  and $n_{l}(z,t)$ for the 
"liquid-like" particles (dashed blue line). 
The interface position $z_{0}(t)$ 
is determined from the crossing point of the thick red and dashed blue lines. 
 \label{fig-1}}
\end{figure}

Using real-space confocal microscopy techniques, Dullens and coworkers \cite{Dullens} 
have recently measured a dynamical 
broadening of the hard sphere solid-fluid interface during sedimentation.
This points to the essential role of crystal nucleation in the supersaturated fluid
and the subsequent built-in of "crystal packages" into the interface.
In a very slowly growing interface, and at low Peclet numbers, 
  a broadening of the fluid-solid interfacial width
relative to its equilibrium value was found. In this letter, we address the broadening of the 
solid-fluid hard sphere interface in the sedimentation process by extensive 
Brownian dynamics computer simulations. We find 
two different types  of broadenings of the fluid-crystal interface: the first occurs
during growth and is correlated with the interface velocity. 
This type of broadening is not driven by gravity as it also shows up in the gravity-free case
{  \cite{footnote}}.
The second type of interface broadening occurs before the splitting of the 
crystal-{ amorphous}  interface from the 
 { amorphous-liquid} interface. The separated interface includes a region which
 is structurally disordered and dynamically arrested.
The latter width behaves nonmonotonic in the gravitational 
driving strength (or Peclet number $Pe$) and exhibits a huge peak relative to its 
equilibrium counterpart. 
While the first type of broadening has not yet
accessed in  confocal microscopy, the latter is consistent with the measurements of 
Dullens et al \cite{Dullens}. However, their measurements were only performed
for low Peclet numbers before the splitting of two interfaces happen. 
\begin{figure}  [!ht]
\begin{center}
\includegraphics*[width=0.9\textwidth]{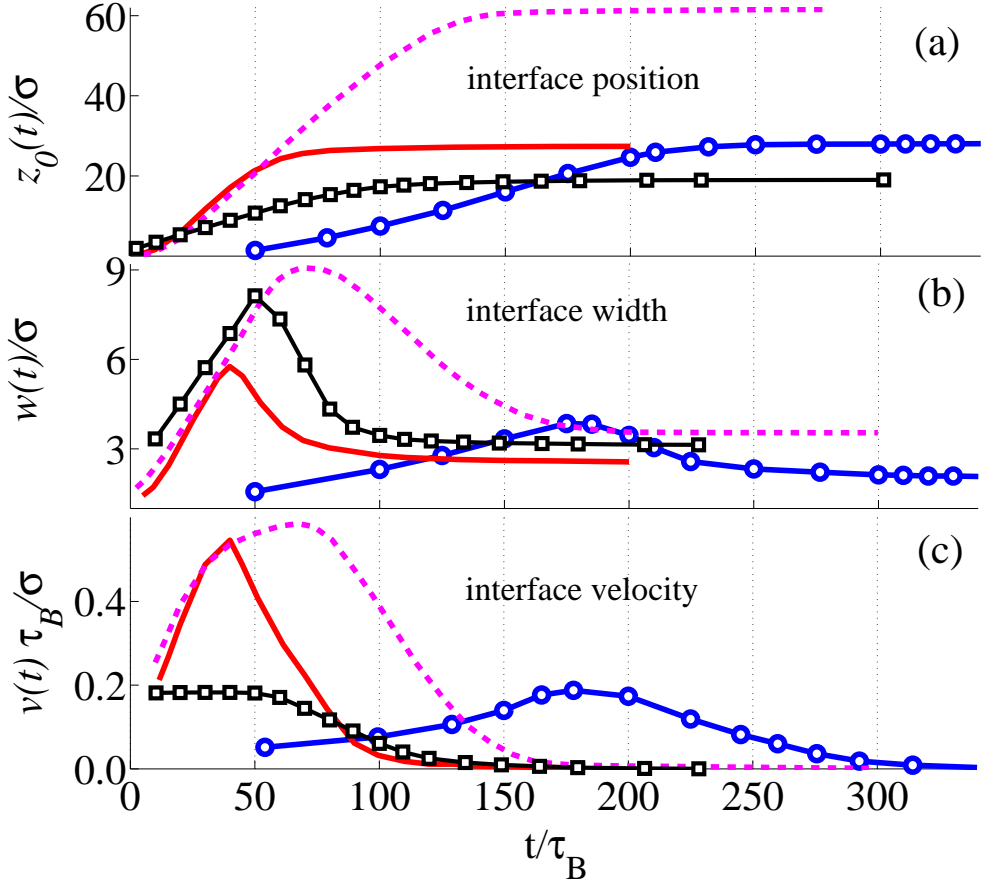}
\end{center}
\caption{Reduced interface position $z_{0}(t)/\sigma$ (a), reduced interfacial width
$w(t)/\sigma$ (b), and reduced propagation velocity $v(t)\tau_B /\sigma$ (c) 
as a function of reduced time $t/\tau_B$. Full blue lines with circles 
are for $\eta_{A}=45.4$ and $\phi=0.1$, 
full red lines are for $\eta_{A}=45.4$ and $\phi=0.3$, dashed pink lines are for 
 $\eta_{A}=90.8$ and $\phi=0.3$. Full black lines with squares are simulation
 data for the  gravity-free ($Pe=0$) suspension with a volume fraction $\phi=0.51$ and $\eta_{A}=46.27$.  
     \label{fig-2}}
\end{figure}

\section{\bf Simulation and Analysis }
In our Brownian dynamics computer simulations, we use an adapted code 
for hard spheres \cite{Cichocki90} of diameter $\sigma$, 
where the short-time infinite-dilution diffusion coefficient $D_0$ sets the Brownian 
time scale $\tau_{\mathrm{B}}=\sigma^2/D_0$. A time step of
 $\Delta t = 0.001\tau_{\mathrm{B}}$ was used in 
integrating the stochastic equations of motion. The simulation box contains 
$N$ hard spheres and has a rectangular shape with dimensions
 $L_{\mathrm{x}} = 40.8\sigma$, 
 $L_{\mathrm{y}} = 43.2\sigma$, and $L_{\mathrm{z}}$ varied between $54\sigma$ and $240\sigma$.  
Various systems with a packing fraction
$\phi=\frac{\pi\sigma^3}{6} \,\frac{N}{L_{\mathrm{x}}L_{\mathrm{y}}L_{\mathrm{z}} }$  
  in the range $0.1 \le \phi \le 0.45$ 
were investigated. 
A further important system parameter is the gravitational load,  or the surface (or areal) density
of particles $\rho_{A}=N/L_{\mathrm{x}}L_{\mathrm{y}}$, which we scale to
\begin{equation}
\eta_{A}=N\sigma^{2}/L_{\mathrm{x}}L_{\mathrm{y}}= \phi {{6 L_{\mathrm{z}}} \over
 {\pi \sigma} }
\end{equation}
Thus different combinations of $\phi$ and $L_{\mathrm{z}}$ result in the same $\eta_A$. 
Simulations were carried for $\eta_A=45.4$ with $N=80\,000$ particles and 
for  $\eta_A=90.8$ with $N= 160\,000$ particles. 
The quantity $\eta_A$ is chosen large enough to produce a crystal-fluid interface 
at a given gravitational strength $Pe$. 
The gravity acceleration $g$ points along the $-z$-direction and has a 
relative strength 
\begin{equation}
Pe = mg\sigma/\left( {2}k_{B}T \right)
% this is the way how me and Dullens define the $Pe$ number.
\end{equation}
where $k_{B}T$ denotes the thermal energy and $m$ the buoyant mass of the colloidal 
particles { \cite{peclet-definition}}.
Periodic boundary 
conditions are employed in the $x$ and $y$ directions, while two hard walls
 are placed at $z=0$ and $z=L_{\mathrm{z}}$.
Next to the wall at $z=0$
 we place a triangular layer of fixed spheres with a lattice constant
 $a = 1.133\sigma$ which acts as an initial template for crystal growth \cite{Heni}.
{ Without a template the crystallization happens 
in a few bottom layers containing small grains, large defects and fault stackings. This 
strongly suppresses the formation and following upward propagation of a single-phase 
crystalline front along the sediment. }
\begin{figure}  [!ht]
\begin{center}
\includegraphics*[width=0.9\textwidth]{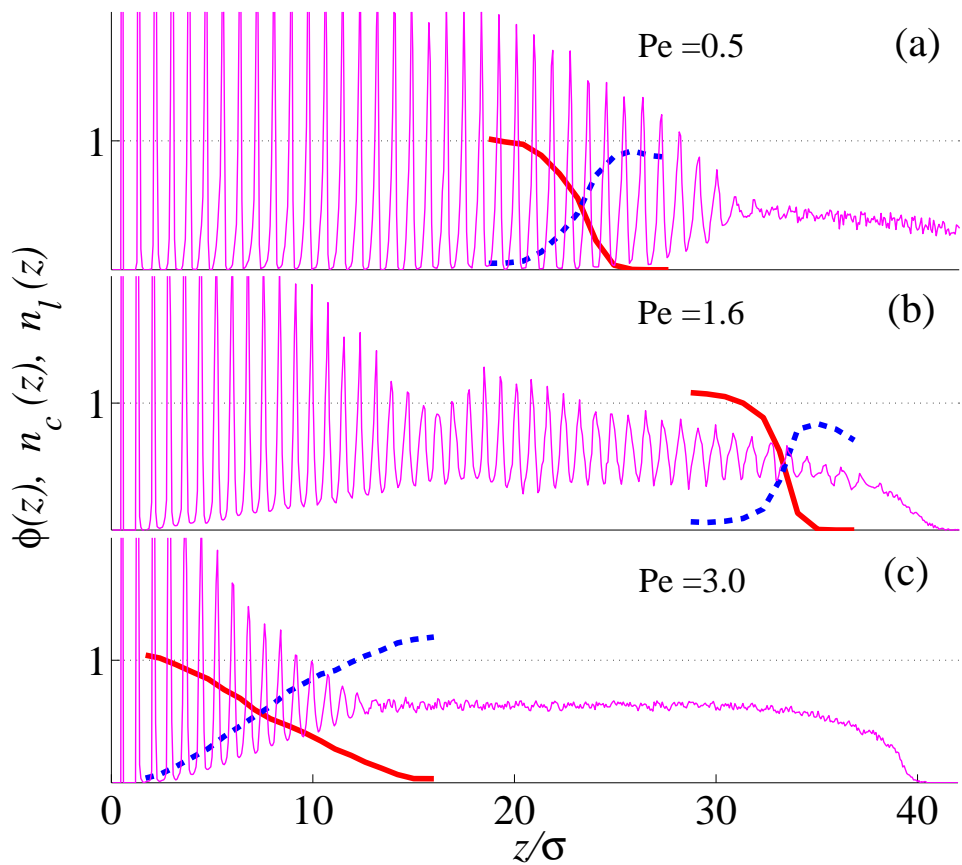}
\end{center}
\caption{Same as Figure~\ref{fig-1}, but now for a fixed large time $t=t_{l}=500\tau_B$
and three different Peclet numbers: (a) $Pe=0.5$, (b) $Pe=1.6$, (c) $Pe=3.0$
at $\eta_{A}=45.4$ and $\phi=0.1$.
 \label{fig-3}}
\end{figure}

All simulations were started from an initial configuration with a homogeneous 
distribution of colloids in the simulation box except the template particles 
in the seed layer next to the bottom wall. This mimics an initially stirred
solution to which gravity is applied instantaneously \cite{Soft_Matter}.
A total simulation time of $500 \tau_{\mathrm{B}}$ was accessed during our simulations.
{ Within this time each particle can sediment about $1000 \sigma Pe$, 
a distance at least 5 times larger  than the characteristic sedimentation length 
$h \approx H \times 0.55/\phi$. Here $H$ is the hight of the sedimented segment above which 
the packing fraction of suspension is $\phi<0.5$.  }
We averaged our { final stage and time-independent}
results over 20 different initial configurations in order to improve the statistics.

For subsequent times $t$, we calculated the laterally 
averaged one-particle packing fraction which is defined as 
\begin{equation}
\phi(z,t)= \frac{\pi}{6}\frac{{\sigma}^3}{L_{\mathrm{x}}L_{\mathrm{y}}} \int \int \ dx \ dy\ \rho(x,y,z,t)  
\end{equation}
and is resolved along the vertical $z$-coordinate. Here $\rho(x,y,z,t)$ is the local 
one-body density of particles at a given time $t$. 
We further identified "crystalline" particles with a crystal-like surrounding
according to a commonly used criterion \cite{Sandomirski}:
 the local orientational-order parameter $\vec q_6(i)$ is
 calculated for each particle $i$ \cite{Steinhardt83,Wolde95}. When two
 particles  $i$ and $j$ are separated by $r\le 1.3\sigma$, we associate
 a crystalline bond to these particles, if  $\vec q_6(i)\cdot \vec q_6(j)>0.5$.
A particle which has  at least 8 of these bonds is considered to be crystalline.
All other particles
are identified as a "liquid-like" particles.
The corresponding local packing fractions 
of crystalline particles $n_{c}(z,t)$ and liquid particles $n_{l}(z,t)$ are also calculated 
and give rise to a $q_{6}$ interface, an example of which is given in 
 Figure~\ref{fig-1}. 
We now define the {\it interface position\/} $z_{0}(t)$ by the implicit condition
$n_{c}(z_{0}(t),t)=n_{l}(z_{0}(t),t)$, i.e. by the position where the fraction of 
crystalline-particles equals that  of the liquid-like particles.
The {\it interfacial width\/} $w(t)$ is defined  as the inverse
 of the maximal slope of the $n_{c}$ profiles
\begin{equation}
w(t) = \left( \partial n_{c}(z,t)\over \partial z \right)^{-1}|_{z=z_{0}(t)}
\end{equation}
We finally define the {\it propagation velocity\/} of the solid-fluid interface as
$v(t)=d z_{0}(t)/ dt$.

\section{\bf Mechanism 1: Kinetic Broadening}
Results for the interfacial profiles 
 at three different times are shown in Figure~\ref{fig-1}. 
Clearly the crystal-fluid interface is propagating from the bottom of the container
 (left side in the figure) into the fluid.
{The lowering of peak heights in the density profile  of  Figure~\ref{fig-1}c
in the region between $z=14\sigma$ and 
 $z=27\sigma$ originates from  the  stacking faults  \cite{Mori2,Mori-2006} 
and defects introduced by a 
small amount of rhcp and hcp crystalline grains. 
These defects tend to dissipate into fcc crystalline layers during longer time simulations.
The system size also plays a crucial role in the dissipation of stacking 
defects \cite{Mori-2011}.} 
When the crystal grows, the interface is getting broader. This is clearly
illustrated in Figure~\ref{fig-2} where the interface position $z_{0}(t)$, its width $w(t)$ 
and its propagation velocity $v(t)$ are shown simultaneously. The interface velocity is 
non-monotonic in time: it first gets accelerated,  reaches a maximum and decays to zero.
 The initial acceleration  has to do with an "induction time" to 
get particles into positions which are structurally favorable for 
subsequent crystal growth. The final
slowing-down is  due to the approach to equilibrium where ideally a finite height
of the interface is reached. Concomitantly with the non-monotonic 
behavior of the interface velocity,
there is a broadening of the interface: as a function of time, 
it first broadens and then shrinks again.
The saturation value at large times, however, is significantly larger  than the initial width.
A direct comparison of Figure~\ref{fig-2}b and \ref{fig-2}c reveals that the maximum 
of the propagation velocity coincides with the maximal interfacial width demonstrating 
that these two phenomena are correlated.
The effect is stable for  different overall densities $\eta_{A}$. 
An increased $\eta_{A}$ just
retards the occurrence of the velocity and width maximum such that they occur at
 larger simulation times. A similar retardation effect takes place in systems with 
low packing fraction $\phi$ at a fixed areal density $\eta_A$, compare solid lines with and without 
symbols in Figure~\ref{fig-2}.   

As a first type of interfacial widening, we 
therefore identify a {\it kinetic broadening\/} of the interface: upon growth,
 the $q_{6}$-interface %flattens
{ becomes less steep},
 see thick lines in Figure~\ref{fig-1}b.
This has to do with the fact that the system has less time to structurally arrange when the interface
propagates quickly. An additional simulation for zero gravity revealed that 
the kinetic broadening effect exists also at zero Peclet number 
{ \cite{gravity-free}}.
However, in the gravity-free case the front growth velocity has no maximum.  
As a reference, simulation data for the
front position, interfacial width and the 
propagation velocity for a coexisting fluid and crystal \cite{Horbach,Mori-2007,Sandomirski} in 
a gravity-free system are included into Figure~\ref{fig-2} as full lines with squares.   
\begin{figure}  [!ht]
\begin{center}
\includegraphics*[width=0.75\textwidth]{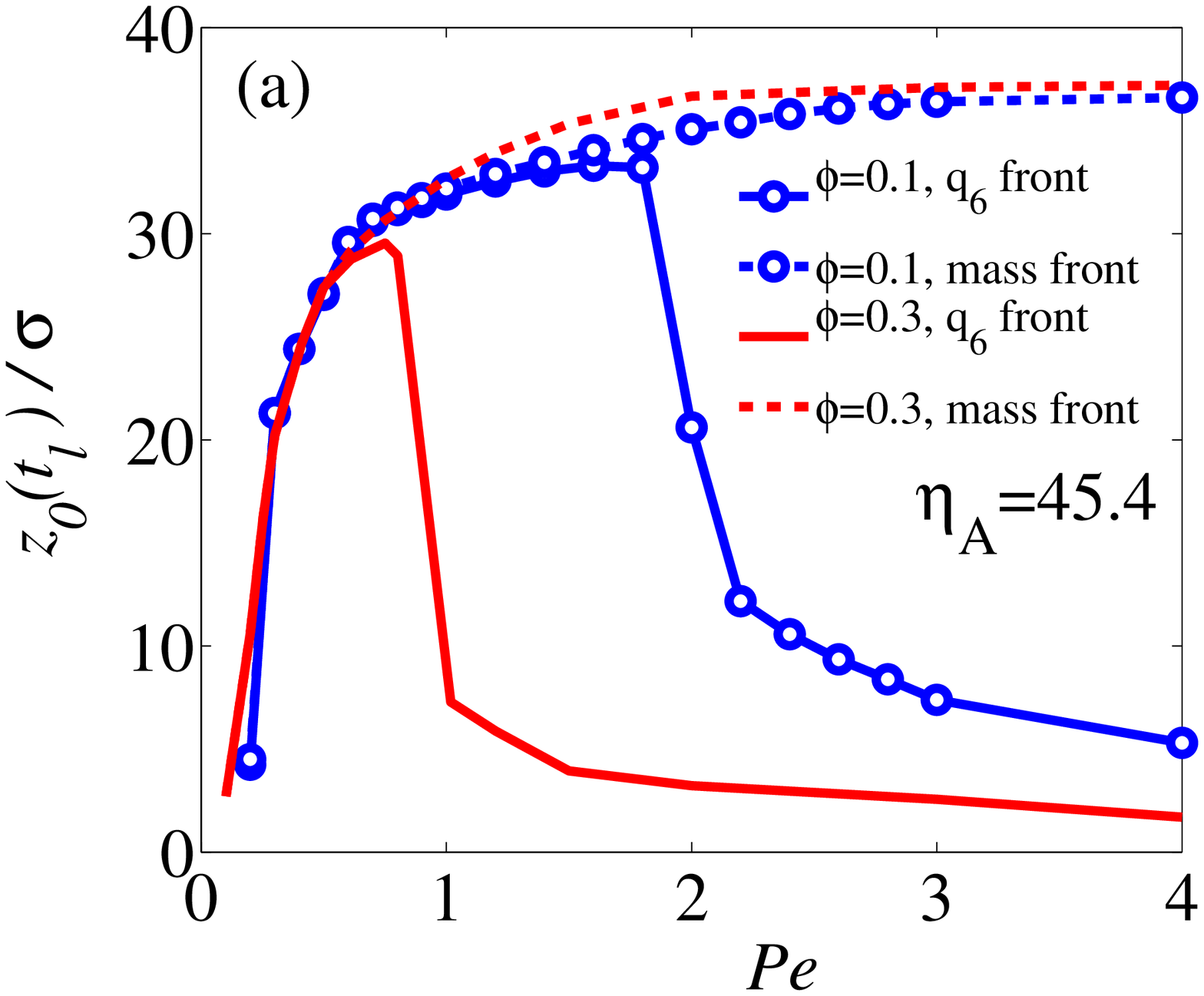}
\includegraphics*[width=0.75\textwidth]{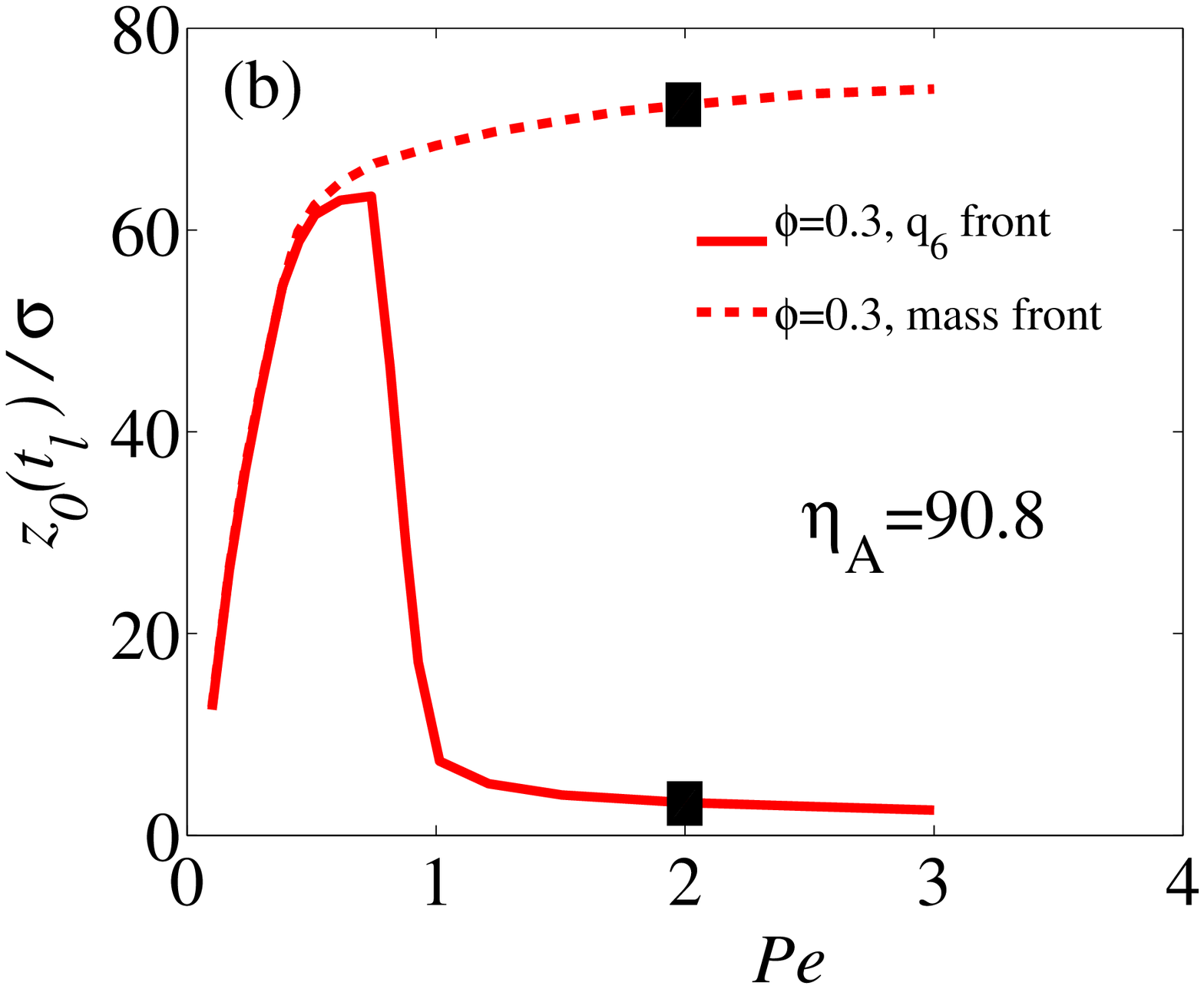}
\end{center}
\caption{(a) Position $z_{0}(t_{l})/\sigma$ of the $q_{6}$-interface (full lines), 
and of the 
{ amorphous-liquid}
 interface (dashed lines) versus Peclet number $Pe$ for two areal densities
 $\eta_{A}=45.4$ (a) and $\eta_{A}=90.8$ (b) at a 
fixed large time $t=t_{l}=500\tau_B$. 
The interface splitting is clearly visible. Black squares at $Pe=2$ correspond to the
{ amorphous-liquid} and 
$q_6$-interfaces analyzed in Figure~\ref{fig-5}.    
  \label{fig-4}}
\end{figure}

\section{\bf Mechanism 2: Interface Splitting}
\begin{figure}  [!ht]
\begin{center}
\includegraphics*[width=1.\textwidth]{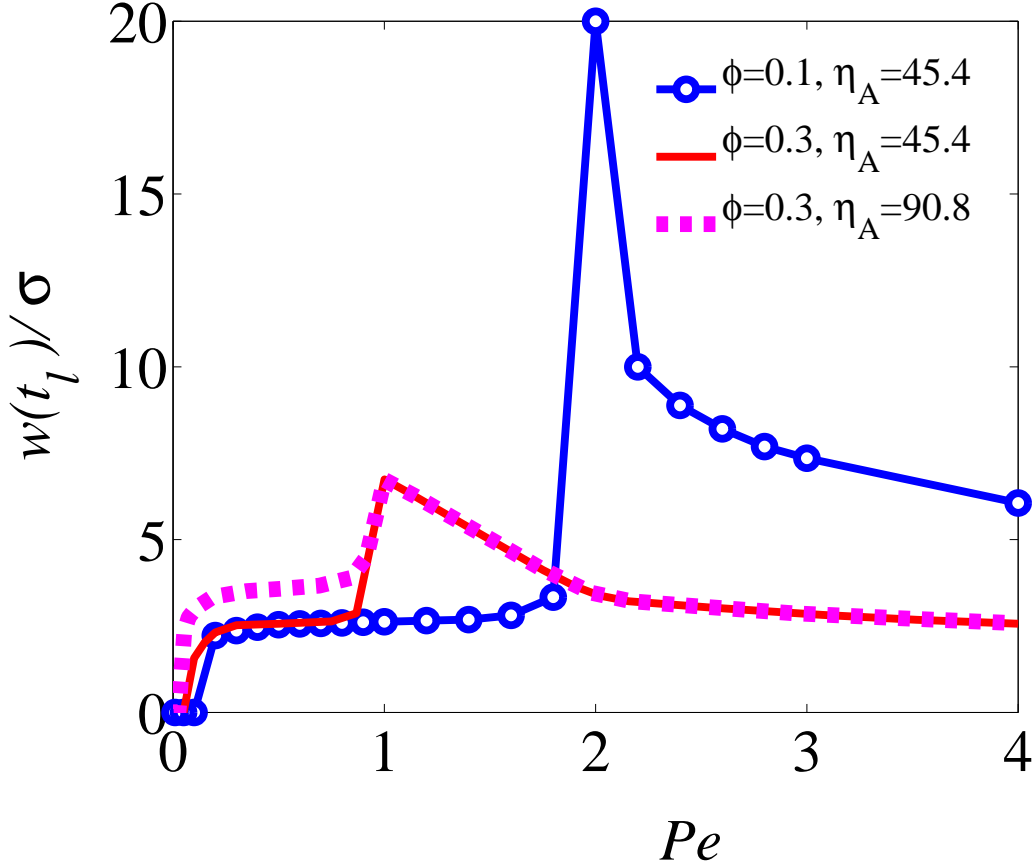}
\end{center}
\caption{ Interfacial width $w(t_{l})/\sigma$ versus Peclet number $Pe$ 
 at a fixed large time $t=t_{l}=500\tau_B$ for two different 
packing fractions $\phi=0.1$ (line with symbols) and 
$\phi=0.3$ (full line) at fixed areal density $\eta_{A}=45.4$, and  
for packing fraction $\phi=0.3$ and $\eta_{A}=90.8$ (dashed line).
 \label{fig-6}}
\end{figure}

We now address the crystal-fluid interface in the final state of the sediment
after a long time $t_{l} = 500 \tau_{\mathrm{B}}$ and discuss its properties as a 
function of the Peclet number (at fixed $\eta_{A}$). In Figure~\ref{fig-3}, 
the laterally-averaged packing fraction profiles
are shown at $t_{l}$ for three different Peclet numbers $Pe=0.5, 1.6, 3.0$.
Interestingly, the position of the interface 
(as visualized by the crossing of the red and blue lines)
behaves non-monotonic in the Peclet number, see also Figure~\ref{fig-4}, which contradicts the 
expectation based on equilibrium arguments. This expectation 
is based on a  discussion of the two limiting 
cases $Pe\to 0$ and $Pe\to \infty$ in equilibrium: for zero gravity, 
the density profile is homogeneously
distributed over the full simulation box (apart from local density correlations at the hard wall)
with a low bulk volume fraction. Hence the solid-fluid interface (if at all)
is close to the bottom wall.
In the opposite limit $Pe\to \infty$, there is a finite number of  layers 
  with a closed packed
density which then abruptly drops to a vacuum at higher $Pe$ at a height of about 
{ $30\sigma$}.
Interpolating between these two limiting cases, in equilibrium, the interface position 
increases with Peclet number $Pe$. 
\begin{figure}  [!ht]
\begin{center}
\includegraphics*[width=0.9\textwidth]{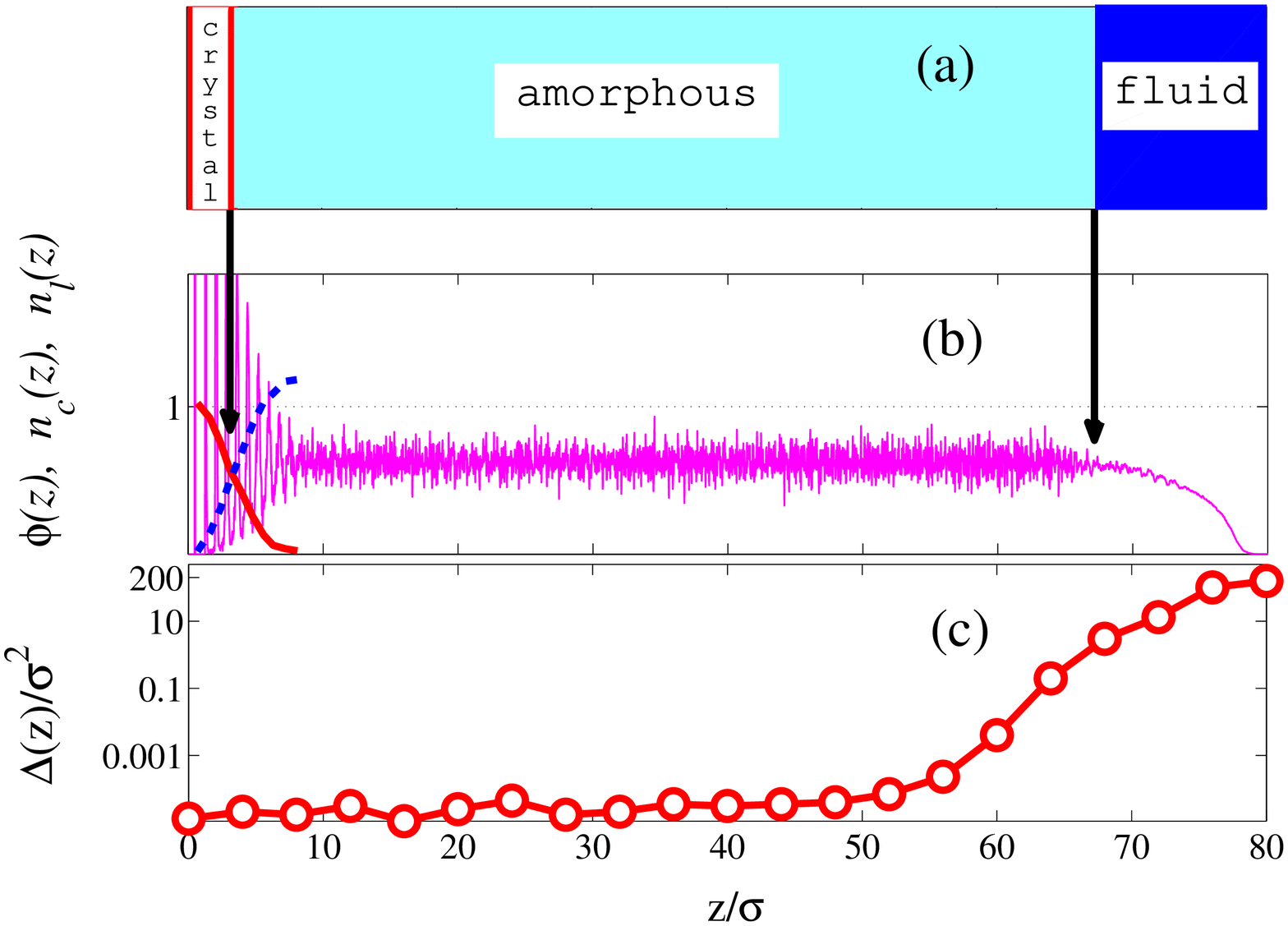}
\end{center}
\caption{ Simulation results for the sediment at $t=t_l=500\tau_B$ for system 
parameters $Pe$=2, $\phi=0.3$ and $\eta_A=90.8$. 
(a) Schematic picture showing the crystalline, amorphous and fluid
zones in the sediment. (b) Laterally averaged packing fraction $\phi(z)$ (thin pink lines), 
 interfacial profile $n_{c}(z)$ for the "crystalline" particles  (thick red line), 
 interfacial profile $n_{l}(z)$ for the
"liquid-like" particles (dashed blue line) versus reduced height $z/\sigma$. 
(c) Mean square displacement  $\Delta(z)$ of the particles along the sediment
obtained within a time window of $\Delta t=10\tau_B$.
  The positions of the crystal-glass and glass-fluid interfaces correspond to the black 
squares in Figure~\ref{fig-4}b.
  \label{fig-5}}
\end{figure}
%---------------------------------------------------------------
{Starting from  $Pe \approx 0.8$ the  single crystalline-liquid interface splits 
into two separate interfaces, the crystalline-amorphous
 and the amorphous-liquid interfaces. 
When the gravitational load increases, the amount of amorphous glass grows 
 partly from the sedimented liquid particles, and partly from the collapsed  
 crystalline layers. Higher loads completely destroy the 
crystal leaving only a few intact layers.}
Besides of the $q_6$-interface position, a position of the
{ amorphous-liquid}
 interface can be defined at the height where the
% full locally averaged 
{ coarse-grained density (averaged over the oscillations)} 
\cite{footnote1} equals the bulk coexisting fluid volume fraction of 
% 0.494
{ 0.492}.
As opposed to the nonmonotonic variation of the $q_{6}$-interface position, 
the position of the 
{ amorphous-liquid} 
 interface is monotonic, see Figure~\ref{fig-4}.
While the positions of the $q_{6}$-interface and the 
{ amorphous-liquid} 
interface coincide for small Peclet number,
 there is a  {\it splitting\/} above a threshold leading to the non-monotonic behavior.
The non-monotonicity can be explained by the formation of a dynamically arrested region 
between the  splitted $q_{6}$ and 
{ amorphous-liquid} 
 interfaces schematically shown in Figure~\ref{fig-5}a
for a system with $\phi=0.3$, $\eta_A=90.8$ and $Pe=2$.  
Within the glassy amorphous region the particles are structurally
disordered, see Figure~\ref{fig-5}b,
 and dynamically caged such that they cannot find suitable surrounding 
to nucleate into a big, and layered, crystal.
This is documented by the smallness of the  
mean square displacement of particles in the amorphous layer, 
see Figure~\ref{fig-5}c. 
The amorphous dynamically arrested part  stops the propagation of the $q_{6}$-interface.
Strong gravity acts therefore like a fast and deep compression, a situation which 
favours glass formation in general. The disordered region does not seem to be composed
of  polycrystalline material as this would have resulted in a finite fraction of crystalline particles.
As shown in Figure~\ref{fig-4}a and \ref{fig-4}b, the Peclet number threshold  for splitting
decreases for increasing packing fraction $\phi$
at fixed $\eta_A$, but does not depend on the imposed overall density $\eta_{A}$
for  fixed $\phi$.
Closely above the Peclet number threshold at which interface splitting occurs, the width $w$
exhibits a strong peak, see Figure~\ref{fig-6}. In fact, the width increases by an order of magnitude 
relative to its low-Peclet-number value and then decreases again. This surprising huge broadening
points to an extremely broad interface between the initial crystalline layers and the 
subsequently formed disordered material. The subsequent decrease for higher Peclet number 
is qualitatively understood by considering the limit $Pe\to \infty$:
 if everything is dominated by strong gravity, the interface then is expected to sharpen 
due to enforced crystallization.

We now compare our results to the real-space experiment by Dullens and coworkers \cite{Dullens}.
The broadening observed in Ref.\ \cite{Dullens} was measured for long times at relatively 
small Peclet numbers (smaller than 0.8) and for larger gravitational loads
$\eta_{A}$ than considered in our work. 
The interfacial velocity was small but nonzero. This implies that the broadening
observed in \cite{Dullens} is consistent with our findings at long times for low $Pe$,
{\it before\/} the interface splitting occurs. It would be interesting to do further 
experiments at higher $Pe$ to verify the interface splitting and the 
huge interface broadening predicted by our simulations.

\section{\bf Conclusion}
In conclusion, we have studied the width of a crystal-fluid interface in a 
sedimenting suspension of hard spheres by Brownian dynamics computer simulations.
Two qualitatively different types of broadening are observed.
The first type of interfacial broadening is purely kinetic. It is not triggered by gravity
but just correlates with the interface velocity. The second type of broadening
is almost static and is huge if the structural crystallinity interface and the
{ amorphous-liquid}
 interface split. 
Between these two interfaces, an amorphous dense sediment is formed which exhibits
a wide structural interface with the lower crystalline part of the sample.
This prediction can in principle be verified by real-space experiments \cite{Dullens,Sandomirski}.
It might also be important for granulates \cite{Hong} where the Peclet number is 
high, and for crystallization in complex plasma \cite{rubin-zuzic}.

Real experimental samples possess an intrinsic polydispersity. 
Furthermore there are solvent-mediated hydrodynamic interactions. 
Most of the influence of hydrodynamic interactions can be captures by an effective 
short-time diffusion constant $D_0$. 
Nevertheless the influence of polydispersity \cite{Sandomirski} and 
hydrodynamic interactions \cite{Soft_Matter,Ladd,Padding} on 
crystal growth remains for future exploration. Furthermore new effects are expected 
for attractive systems \cite{Leocmach} as witnessed by different dynamically arrested states
(attractive glasses) \cite{Zaccarelli} which could impede crystallization during sedimentation.
The latter case is realized for colloid-polymer mixtures.

{It is worth to mention that some of the present findings on the interface broadenings   
 may be characteristics of fcc \{111\} growth from a patterned template with triangular 
lattice considered in current study. It will be useful to check whether 
a similar  broadenings takes place  in the \{100\} growth, a case considered 
in Refs.~\cite{Horbach,Mori-2007} for gravitation-free hard sphere systems.  
}

In general, our findings can be used to steer the thickness of crystalline and amorphous
layers as well as their interfacial structure by sedimentation. This may help
to open the way for designing smart colloidal materials with new optical
and rheological properties.

\acknowledgments
We thank R. Dullens, M. Marechal, S. U. Egelhaaf and K. Sandormirski for helpful discussions.
Financial support from the DFG within SPP 1296 is gratefully acknowledged.
E.A. also acknowledges partial support of this work by the US Department of Energy under grant
DE-FG02-05ER46244.


\begin{thebibliography}{0}
\bibitem{peclet-definition} 
{
In colloidal physics and fluid mechanics the motion of a solute particle under external driving field  
is characterized by a Peclet number $Pe=\tau_D/\tau_S$, which represents the
competition between advection and diffusion processes.
  Here $\tau_D$ is the time it takes
a particle to diffuse its own radius $\tau_D= \sigma^2/(4D_0)$, 
 and $\tau_S$ is the time it takes the particle to sediment the same 
distance $\tau_S = \sigma/(2 \upsilon_S)$.
 Taking into account the fact that the sedimentation velocity $\upsilon_S$ of the solute 
 particle can be approximated by its terminal velocity 
under the force of gravitational drag, $\upsilon_S =mg/\xi$, and the diffusion coefficient of 
the solute particle is $D_0=k_BT/\xi$ ($\xi$ is the Stokes friction coefficient), 
the Peclet number can be rewritten as: 
$Pe={{\sigma}\over{2}} {{\upsilon_S}\over {D_0}} =
 {{\sigma} \over {2}} {\frac {(mg/\xi)} {(k_BT/\xi)}} = 
 {{\sigma} \over {2}}  \frac {mg} {k_BT}$.  
}
\bibitem{footnote1} In detail, the 
%fully averaged 
{ coarse-grained}
density is obtained from the $z$-resolved profiles
by integrating between two subsequent minima.
\end{thebibliography}
\end{document}